\documentclass[english]{article}

\usepackage{geometry} 
\usepackage{graphicx}
\usepackage{amssymb}
\usepackage{epstopdf}
\usepackage{babel}
\usepackage{inputenc}
\usepackage{epsfig}

\def\aprle{\buildrel < \over {_{\sim}}}

\begin{document}

\vspace{0.8cm}

\begin{center}
{\Large {\bf Magnetic Monopole Search at high altitude with the SLIM 
experiment}} 

\vspace{0.8cm}

\normalsize{
S. Balestra$^{1,2}$, 
S. Cecchini$^{1,3}$,
M. Cozzi$^{1,2}$,
M. Errico$^{1,2}$,
F. Fabbri$^2$, 
G. Giacomelli$^{1,2}$, 
R. Giacomelli$^2$, 
M. Giorgini$^{1,2}$, 
A. Kumar$^{1,4}$, 
S. Manzoor$^{1,5}$, 
J. McDonald$^6$,
G. Mandrioli$^2$, 
S. Marcellini$^2$,
A. Margiotta$^{1,2}$, 
E.~ Medinaceli$^{1,7}$,  
L. Patrizii$^2$, 
J. Pinfold$^6$, 
V. Popa$^{2,8}$, 
I.E. Qureshi$^5$,
O. Saavedra$^{9,10}$,
Z. Sahnoun$^{2,11}$, 
G. Sirri$^2$,  
M. Spurio$^{1,2}$, 
V. Togo$^2$, 
A. Velarde$^7$ and
A. Zanini$^{10}$

\par~\par

{\small\it
(1) Dip. Fisica dell'Universit\'a di Bologna, 40127 Bologna, 
Italy \\  
(2) INFN Sez. Bologna, 40127 Bologna, Italy\\
(3) INAF/IASF Sez. Bologna, 40129 Bologna, Italy\\
(4) Physics Dept., Sant Longowal Institute of Eng. \& Tech., Longowal, 
148 106, India\\  
(5) PD, PINSTECH, P.O. Nilore, and COMSATS-CIIT, No. 30, H-8/1, Islamabad, Pakistan\\ 
(6) Centre for Subatomic Research, Univ. of Alberta, Edmonton, 
Alberta T6G 2N4, Canada\\ 
(7) Laboratorio de F\'{i}sica C\'{o}smica de Chacaltaya, UMSA, La Paz, Bolivia\\ 
(8) Institute for Space Sciences, 077125 Bucharest-M\u{a}gurele, Romania\\
(9) Dip. Fisica Sperimentale e Generale, Universit\'a di 
Torino, 10125 Torino, Italy\\ 
(10) INFN Sez. Torino, 10125 Torino, Italy\\
(11) Astrophysics Dept., CRAAG, BP 63 Bouzareah, 16340 Algiers, Algeria}
}

\end{center}

\vspace{1cm}

\begin{center}
{\bf Abstract} 
\end{center}

{\normalsize 
The SLIM experiment was a large array of
nuclear track detectors located at the Chacaltaya high altitude
Laboratory (5230 m a.s.l.). The detector was in particular sensitive to 
Intermediate Mass Magnetic Monopoles, with masses $10^5<M_M<10^{12}$ GeV. From 
the analysis of
the full detector exposed for more than 4 years a flux upper limit of 
$1.3 \cdot 10^{-15}$ cm$^{-2}$ s$^{-1}$ sr$^{-1}$ for downgoing fast 
Intermediate Mass Monopoles was established at the 90\% C.L. 

\section{Introduction}
\label{intro}
In 1931 Dirac introduced Magnetic Monopoles (MMs) in order to explain the
quantization of the electric charge, obtaining the formula
$eg=n \hbar c/2$, from which $g=ng_D=n \hbar c/2e=n~68.5e=n~3.29 
\cdot 10^{-8}$ in the c.g.s. symmetric system of units \cite{Dirac}; $n$ 
is an integer, $n = 1,2,3,...$ MMs possessing an electric charge and 
bound systems of a magnetic monopole with an atomic nucleus are
called dyons. An extensive bibliography on MMs is given in ref. \cite{biblio}. 
Relatively low mass classical Dirac monopoles have been searched for at 
high energy accelerators \cite{bertani,bakari}.

 Magnetic Monopoles are present in a variety of unified gauge models with 
a wide range of masses. 

 Grand Unified Theories (GUT) of the strong and 
electroweak interactions at the mass scale $M_G \sim 10^{14}\div10^{15}$
GeV predict the existence of magnetic monopoles, produced  
in the early Universe at the end of the GUT epoch, with 
very large masses, $M_M \geq 10^{16}$ GeV. Such monopoles cannot be 
produced with existing accelerators, nor with any foreseen for the 
future. In the past, GUT poles were searched for in the cosmic 
radiation. These poles are characterized by low velocities and 
relatively large energy losses \cite{MMs}. The MACRO
experiment set the best limits on GUT MMs with $g=g_D,~2g_D,~3g_D$ and dyons 
at the level of $\sim 1.4 \cdot 10^{-16}$ cm$^{-2}$ s$^{-1}$ sr$^{-1}$ for 
$4 \cdot 10^{-5} < \beta=v/c < 0.7$ \cite{MACRO}.

Some GUT models and some supersymmetric models predict 
Intermediate Mass Monopoles (IMMs)
with masses $10^{5}< M_{M} < 10^{12}$ GeV and with magnetic charges of 
multiples of $g_D$; these MMs may have been
produced in later phase transitions in the early Universe and could be 
present in the cosmic radiation \cite{IMMs,UHECR}. 

 IMMs may be relativistic since they could be accelerated to high
velocities in one coherent domain of the galactic magnetic field. In
this case one would have to look for downgoing, fast ($\beta> 0.03$), 
 heavily ionizing MMs~\footnote{The interest in MMs was also connected 
with the possibility that they could yield the highest energy cosmic 
rays \cite{UHECR}.}.

The main purpose of the SLIM  (Search for LIght Monopoles) 
experiment at the Chacaltaya laboratory in Bolivia at 5230 m a.s.l., 
 was the search for IMMs \cite{proposal}. An exposure at a high altitude 
laboratory allows to search for MMs of lower masses, higher magnetic 
charges and lower velocities, see Fig. \ref{fig:betavsmass}.

The searches for IMMs by Earth based detectors are essentially limited to 
downgoing particles \cite{1998}. Water Cherenkov detectors are limited to 
fast downgoing IMMs (with $\beta > 0.5$), and a search can be done if the 
detectors are able to discriminate against the large background of cosmic 
ray muons \cite{2007}.

\begin{figure}[!h]
\begin{center}
{\centering\resizebox*{!}{8cm}{\includegraphics{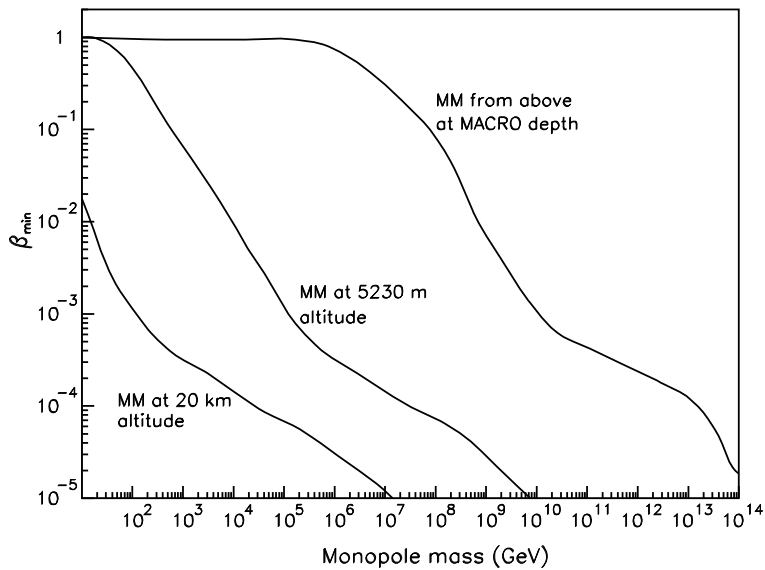}}}
\begin{quote} 
\caption{\small 
Accessible regions (above lines) in the plane (mass, $\beta$) for 
monopoles with magnetic charge $g=g_D$ coming from above for an experiment 
at altitudes of 20000 m, 5230 m,  and for an underground detector at 
the Gran Sasso Lab. (average rock overburden of 3700 m.w.e.)}
\label {fig:betavsmass}
\end{quote}
\end{center}
 \end{figure}

The SLIM detector was also sensitive to Strange Quark Matter 
nuggets \cite{nuclr,SLIM05/5} and Q-balls \cite{qballs}. The results 
on these Dark Matter candidates are discussed in ref. \cite{SQM}. \par
In the following, we present a short description of the SLIM apparatus, 
 the calibrations of the Nuclear Track Detectors (NTDs), the etching and 
analysis procedures, and the limits obtained by the experiment on IMMs 
and GUT Magnetic Monopoles.

\begin{figure}[h!]
 \centering
{\centering\resizebox*{!}{6.5cm}{\includegraphics{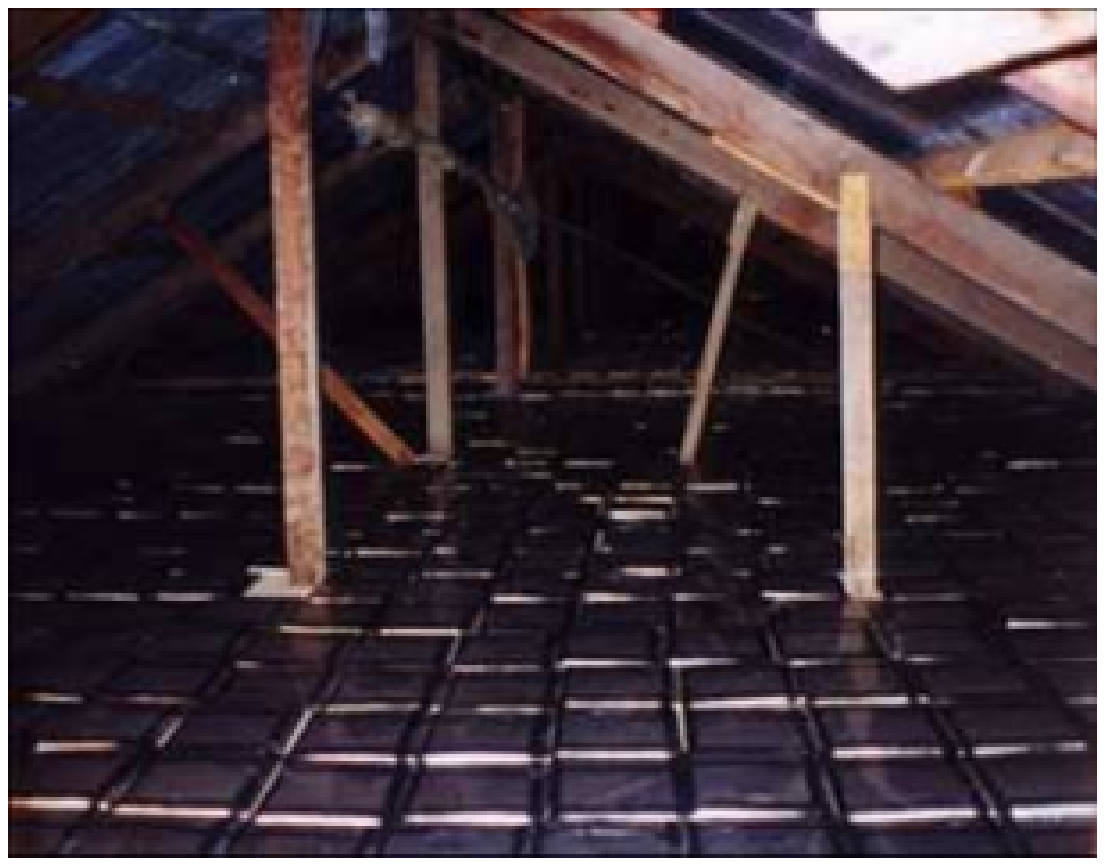}}}
 \hspace{1cm}
 {\centering\resizebox*{!}{6.5cm}{\includegraphics{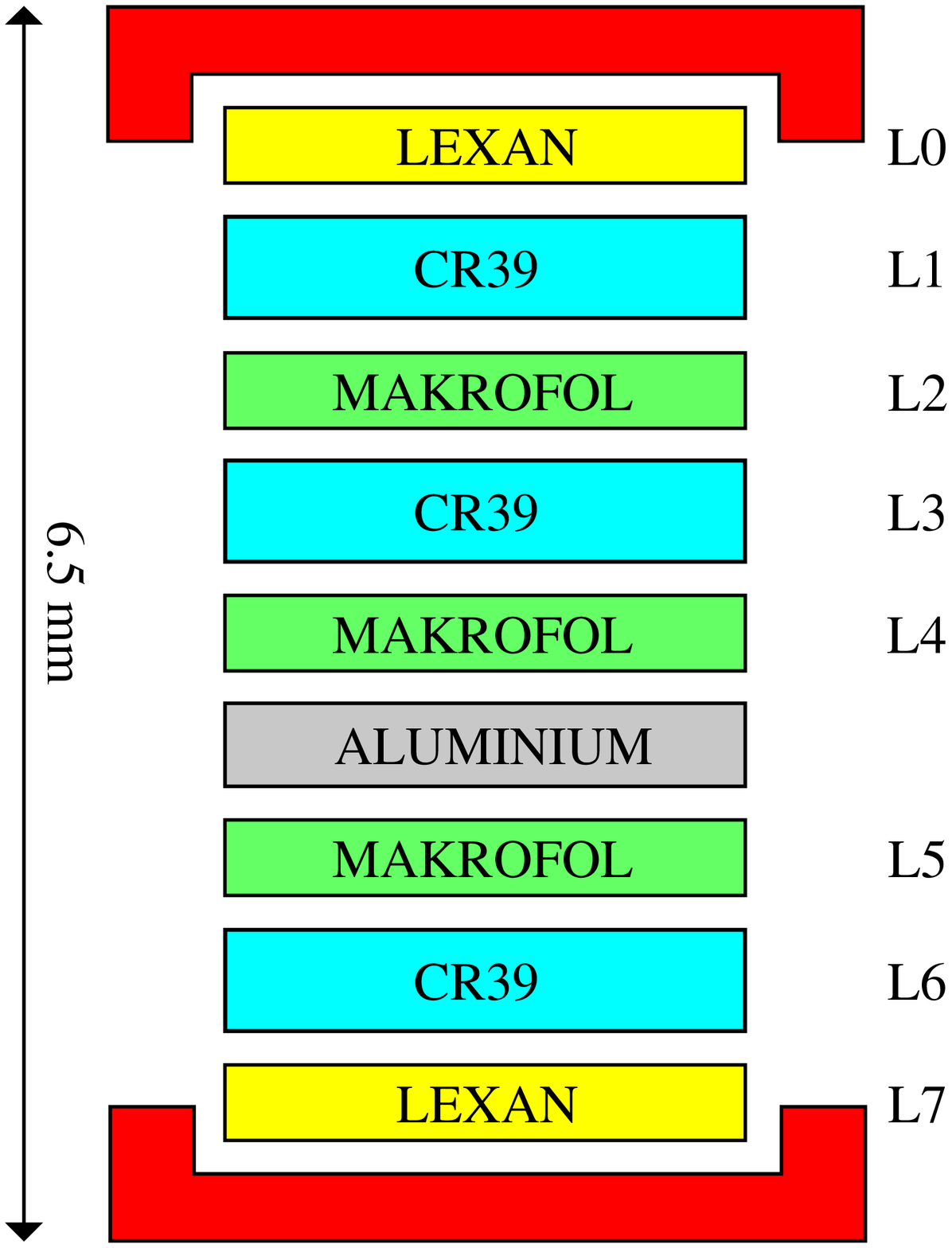}}}
\begin{quote}
 \caption{\small Left: the SLIM modules installed at Chacaltaya. 
Right: composition of one of the 7410 modules; each module was enclosed in an 
aluminized mylar bag filled with dry air at a pressure of 1 bar.} 
\label{fig:illustration2}
\end{quote}
 \end{figure}

\section{Experimental procedure}
\label{sec:experimental}
The SLIM  experiment was an array of NTDs~\footnote{Another
100 m$^2$ of NTDs were installed at Koksil (Pakistan, 4275 m a.s.l.) since
2002 and were not used in the present analysis.} with a total surface area 
slightly greater 
than 400 m$^2$ \cite{proposal}. The array was organized into 7410 
modules, each of area 
$24 \times 24$ cm$^2$. All modules were made up of: three layers of 
CR39$^{\scriptsize \textregistered}$~\footnote{The SLIM CR39 was produced 
by the Intercast Europe Co, Parma, Italy according to our 
specifications.}, each 1.4 mm thick; 3 layers of Makrofol 
DE$^{\scriptsize \textregistered}$~\footnote{Manufactured by Bayer 
AG, Leverkusen, Germany.}, each 0.48 mm thick; 2 layers of Lexan each 
0.25 mm thick and one layer of aluminum absorber 1 mm thick (see Fig. 
\ref{fig:illustration2} right). 
The CR39 used in about 90\% of the modules (377 m$^2$) was of the same type 
used in the MACRO experiment \cite{MACRO}. The remaining modules, 50 
m$^2$, utilized CR39 containing 0.1\% of DOP additive, CR39(DOP).

Each module (stack) was sealed in an aluminized plastic bag (125 $\mu$m 
thick) filled with dry air at a pressure of 1 bar.
The  modules were transported to La Paz, Bolivia, from Italy in wooden 
boxes and their position with respect to the other modules in the 
shipping crate was recorded.  The stacks were deployed under the roof of the
Chacaltaya Laboratory, roughly 4 m above ground (see Fig. 
\ref{fig:illustration2} left). 
The installation of the SLIM detectors started in February 2000 and ended 
in February 2002. The return of the material to Italy was organized in 
batches, after the completion of the 4 years exposure.

The atmospheric pressure at Chacaltaya is about 0.5 bar; before shipping 
to Chacaltaya, in Bologna we checked the air tightness of the envelopes 
sealed with air at a pressure of 1 bar by placing a sample of them 
in an airtight tank at a pressure of 0.3 atm for a few months; no significant 
leakage was detected.
   
From the experience gained with the MACRO Nuclear Track 
Subdetector \cite{MACRO}, we know that the used CR39 does not suffer from 
``aging" or ``fading" effects for exposure times as long as 10 
years \cite{NTDsM}. Further calibrations with 1 AGeV Fe$^{26+}$ ions in 
1999 and 2005 and with 158 AGeV In$^{49+}$ in 2003 confirmed the quality and 
the stability of the CR39 used in the SLIM experiment \cite{calibrations}.

\subsection{Environmental measurements}
\label{sub:env}     
During the first phases of the detector deployment we evaluated possible  
effects of climatic conditions on the detector response and possible 
backgrounds. Previous tests had shown that the CR39 response does not 
depend on the time elapsed from its production and the passage of the 
particle if the ambient temperature ranges between -20 $^\circ$C and +30 
$^\circ$C. The minimum and maximum values of the air temperature in each 
detector hall in Chacaltaya was recorded 3 times a day over the lifetime 
of the experiment. The temperature values usually ranged from 0 $^\circ$C to 
30 $^\circ$C with an average value of 12 $^\circ$C for the whole year 
and from one year to the other; however in the summer months in very few cases
temperatures down to -5 $^\circ$C were measured in the early morning. 
 Therefore, no significant variations 
were expected in the detector response over the exposure period.

We performed measurements of the radon concentration in different locations 
of the experimental rooms where the SLIM detectors were placed. We used 
for this purpose E-PERM$^{\scriptsize \textregistered}$ radon dosimeters. The 
measured radon activity was about $40 \div 50$ Bq/m$^3$ of air. According 
to our previous experience with the MACRO  NTDs, we concluded that this 
level of radon induced radioactivity did not present a problem for the 
experiment, even in case of radon diffusion into the module bags.

Two different types of neutron detectors (BTI bubble counters and a BF3 
counter detectors) were used to measure the neutron flux at Chacaltaya, during 
the first installation shift of 2001 over the energy range of a few hundred 
keV to about 20 MeV \cite{neutron}. Neutrons of these energies interacting 
inside the detectors could induce  background tracks, and their density 
could affect the scanning speed and efficiency. Both types of neutron 
detectors measured the accumulated dose. Consistent results were obtained 
by both types of detectors. The accumulated dose measured in open air and 
near the detectors was very similar. The absolute neutron flux was computed 
using the BTI bubble counters for which the efficiency is known. A value 
of $(1.7 \pm 0.8) \cdot 10^{-2}$ cm$^{-2}$ s$^{-1}$ was obtained, which is 
in agreement with other reported neutron flux data at the altitude of 
Chacaltaya and with more recent measurements at the same location 
\cite{neutroni}. 
The necessity to reduce the neutron induced background in CR39 required us 
to study special etching procedures, mainly based on the addition of ethyl 
alcohol to the etching solutions. As discussed in the next section, the 
addition of alcohol reduces the background tracks on the detector sheets 
and improves the surface quality (i.e. greater transparency), at the 
expense of a higher threshold \cite{calibrations}.

\subsection{Etching procedures}
\label{sub:etching}
The passage of a magnetic monopole in NTDs, such as CR39, is expected to 
cause structural line damage in the polymer (forming the so called 
``latent track''). Since IMMs have a constant energy loss through the 
stacks, the subsequent chemical etching should result in collinear 
etch-pit cones of equal size on both faces of each detector sheet. 
 In order to increase the  detector ``signal to noise'' ratio different 
etching conditions \cite{NTDsM,calibrations} were defined.
 The so-called ``strong etching'' technique allows better surface quality 
and larger post-etched cones to be obtained. This makes etch pits easier 
to detect under visual scanning. Strong etching was used to analyze the 
top-most  CR39 sheet in each module.
``Soft etching''  was applied to the other CR39 layers in a module if a 
candidate track was found after the first scan. This process allows to 
proceed in several etching steps and study the formation of the 
post-etched cones.

 For CR39 and CR39(DOP) the strong etching conditions were: 8N KOH + 
1.5\% ethyl alcohol at 75~$^\circ$C for 30 hours. The bulk etching velocities 
were $v_B = 7.2~\pm~0.4~ \mu$m/h and $v_B = 5.9 \pm 0.3~ \mu$m/h for 
CR39 and CR39(DOP), respectively.

\begin{figure}[t]
 \centering
{\centering\resizebox*{11cm}{11cm}{\includegraphics{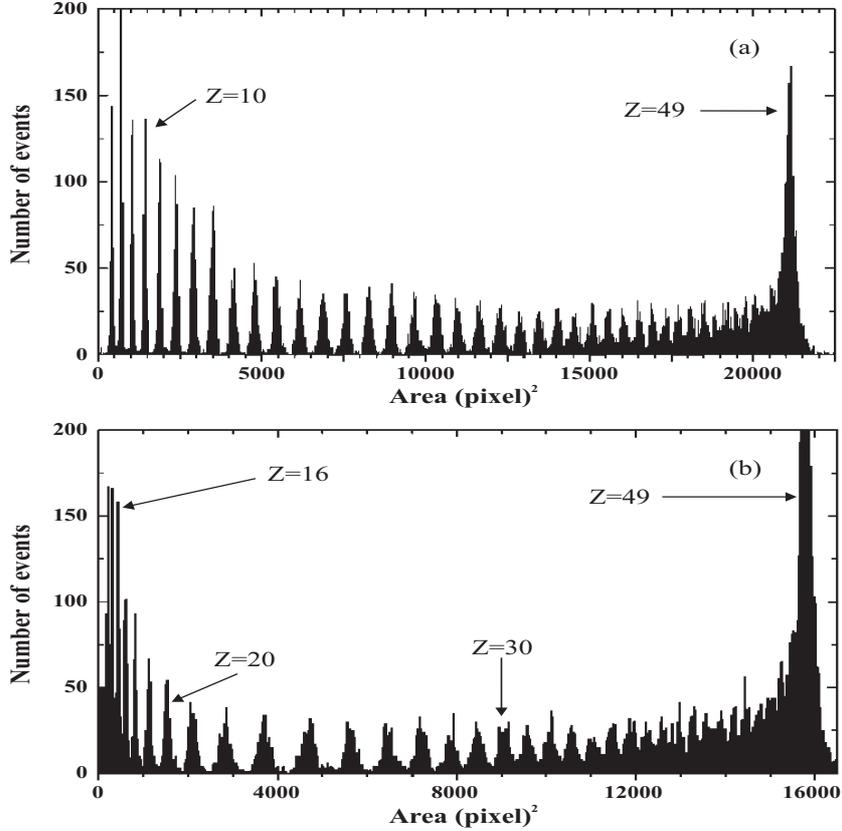}}}
\begin{quote}
 \caption{\small Calibrations of CR39 nuclear track detectors
 with 158 A GeV In$^{49+}$ ions and their nuclear fragments with 
 decreasing charge. The base areas ($1~pixel^2 = 0.3~\mu$m$^2$) of the 
etched cones were averages over 2 faces. The CR39 was etched in (a) soft and 
(b) strong etching conditions.} 
\label{fig:picchi}
\end{quote}
 \end{figure}

The soft etching conditions were 6N NaOH + 1\% ethyl alcohol at 70~$^\circ$C 
for 40 hours for CR39 and 
CR39(DOP). The bulk etching rates were $v_B = 1.25 
\pm 0.02~\mu$m/h and $v_B = 0.98 \pm 0.02~\mu$m/h for CR39 and 
CR39(DOP), respectively.
  
Makrofol NTDs were etched in 6N KOH + 20\% ethyl alcohol at 
50~$^\circ$C for 10 hours; the bulk etch velocity was $v_B = 3.4~\mu$m/h.

\begin{figure}[t]
 \centering
{\centering\resizebox*{!}{5.5cm}{\includegraphics{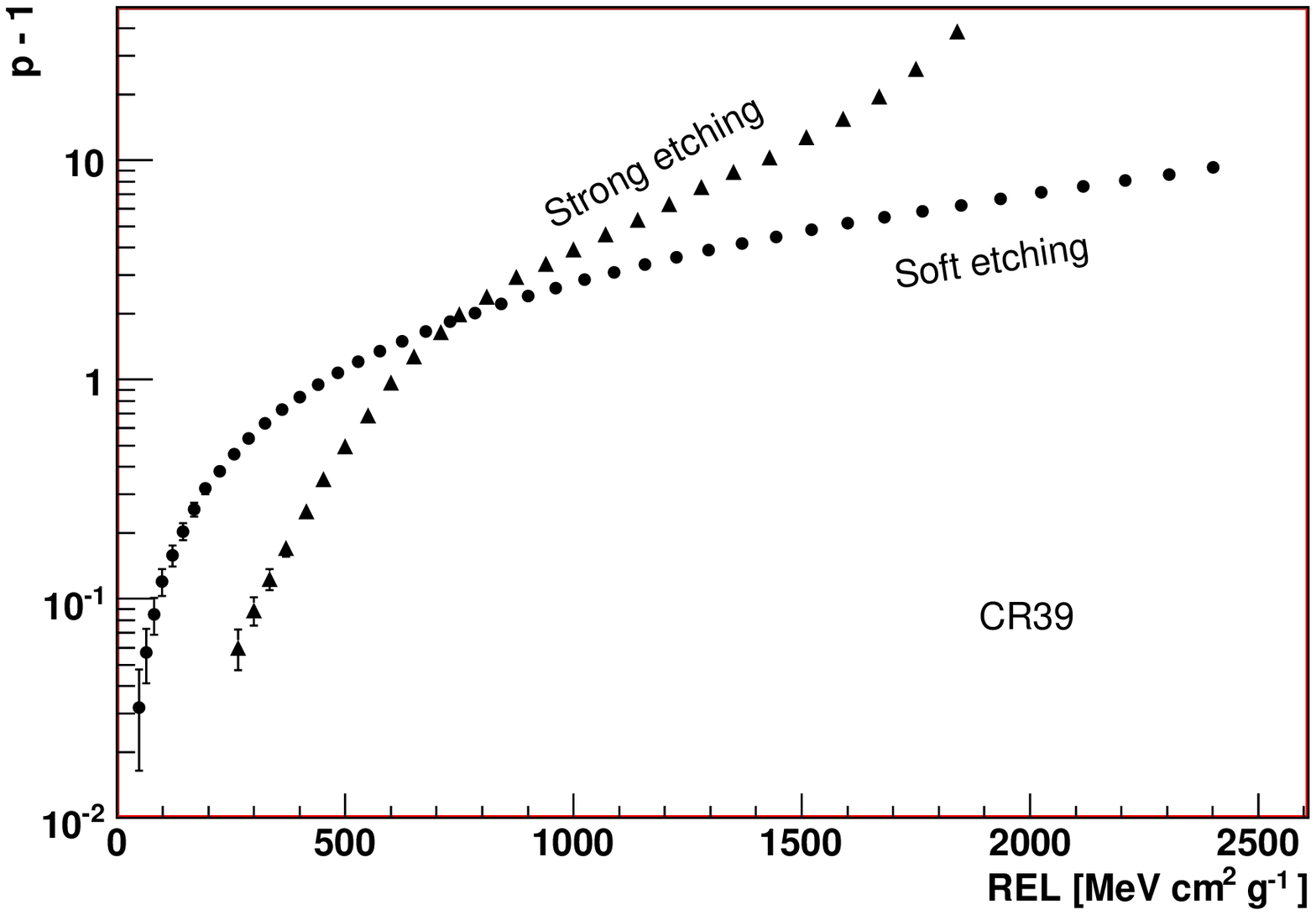}}}
{\centering\resizebox*{!}{5.5cm}{\includegraphics{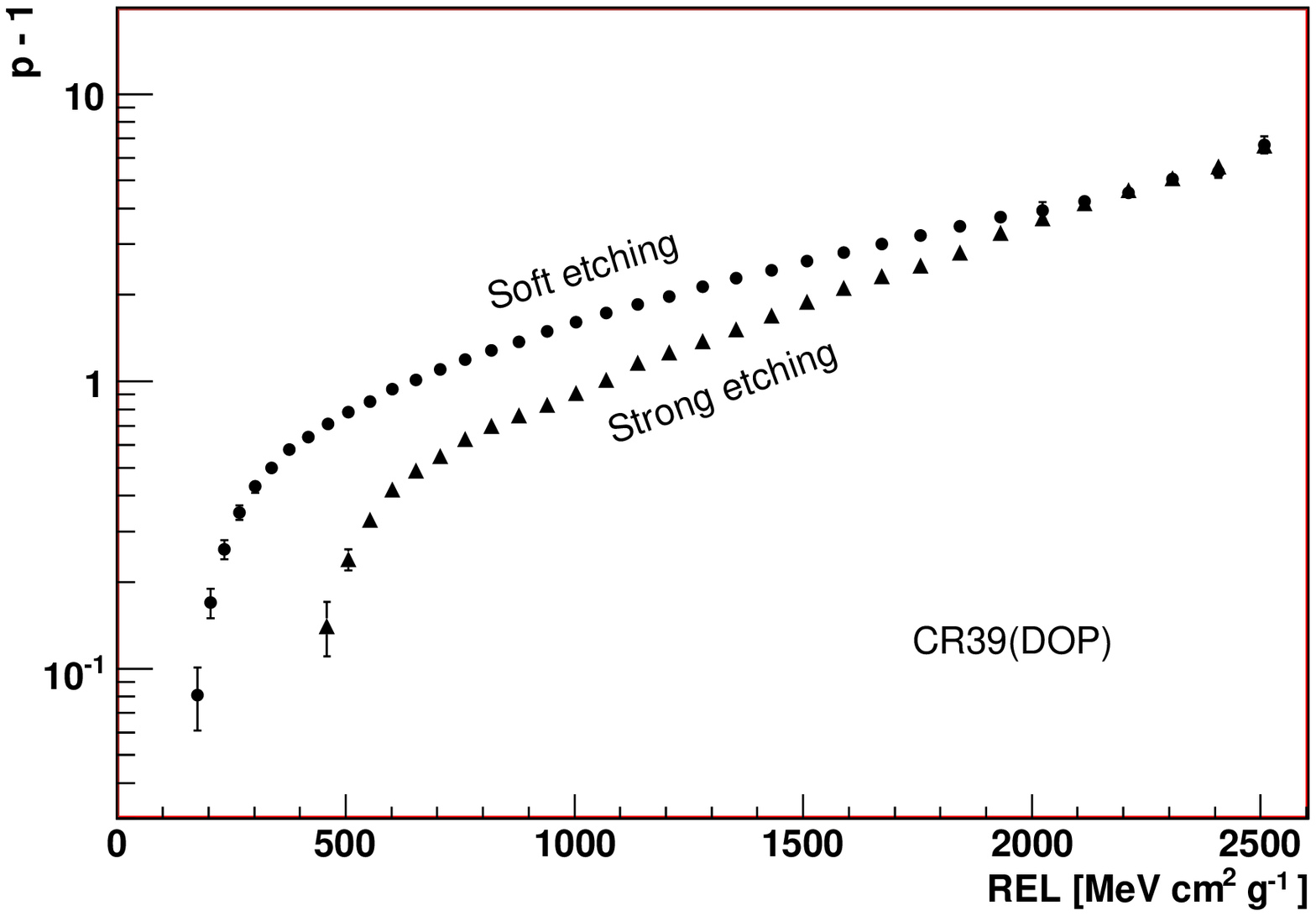}}}
\begin{quote}
 \caption{\small Reduced track etch rate $(p-1)$ vs REL for the 
CR39 (left) and CR39(DOP) (right) detectors, exposed to the 158 A GeV 
indium ion beam, etched in soft and strong etching conditions.} 
\label{fig:tracketch}
\end{quote}
 \end{figure}

\subsection{NTD calibrations}
\label{sub:cal} 
The CR39 and Makrofol nuclear track detectors were calibrated with 158 A GeV 
In$^{49+}$ and Pb$^{82+}$ beams at 
the CERN SPS and 1 A GeV Fe$^{26+}$ at the Brookhaven National Laboratory 
(BNL) Alternating Gradient Synchrotron (AGS). The calibration layout was 
a standard one with a fragmentation target and CR39 (plus Makrofol) NTDs 
in front of and behind the target \cite{fraggg}. 
The detector sheets behind the target detected both primary ions and nuclear 
fragments of decreasing charge.
 
We recall that  the formation of etch-pit cones (``tracks'') in NTDs is 
regulated by the bulk etching rate, $v_{B}$,  and the track etching 
rate, $v_{T}$, i.e. the velocities at which the undamaged and 
 damaged materials (along the particle trajectory), are etched out.  
Etch-pit cones are formed if $v_T > v_B$.
 The response of the CR39 detector is measured by the etching rate ratio 
 $p=v_T / v_B$. 

After etching the standard calibration procedure was the following: 

$(i)$ measure the base area of each track in NTDs with an automatic image 
analyzer system \cite{Elbeck}. The projectile fragments carry the same
$\beta$ and approximately the same direction of the incident ion; the $Z$
of each resolved peak is identified via the base area spectrum. 
 The average base area distributions of the 
In$^{49+}$ ions and of their fragments in CR39, etched in soft or 
 strong conditions, are shown in Figs. \ref{fig:picchi}a,b 
($1~pixel^2 = 0.3~\mu$m$^2$). 

$(ii)$ For each calibration peak the $Z/ \beta$ is obtained 
and the reduced etch rate $(p-1)$ is computed. The Restricted Energy Loss 
(REL) due to ionization 
and nuclear scattering is evaluated, thus arriving to the calibration 
data of $(p-1)$ vs REL shown in Fig. 
\ref{fig:tracketch} for both strong and soft etching conditions for CR39 
and CR39(DOP). For soft etching the threshold in CR39 is at $Z/\beta \sim 7$ 
corresponding to REL $\sim 50$ MeV cm$^2$ g$^{-1}$. For strong etching the 
threshold is at $Z/\beta \sim 14$, corresponding to REL $\sim 200$ MeV 
cm$^2$ g$^{-1}$. The extrapolation of the calibration curves to $p=1$ 
gives REL $\aprle 40$ MeV cm$^2$ g$^{-1}$ for soft etching and REL 
$\aprle 160$ MeV cm$^2$ g$^{-1}$ for strong etching.
 For CR39(DOP) the threshold in soft etching conditions is at $Z/\beta 
\sim 13$ corresponding to REL $\sim 170$ MeV cm$^2$ g$^{-1}$; the threshold 
in strong etching conditions is at $Z/\beta \sim 21$ corresponding to REL 
$\sim 460$ MeV cm$^2$ g$^{-1}$. The extrapolation of the calibration curves 
to $p=1$ gives REL $\aprle 240$ MeV cm$^2$ g$^{-1}$ for strong etching.

For magnetic monopoles with $g=g_D,~2g_D,~3g_D$ we computed the REL as
a function of $\beta$ taking into account electronic and nuclear energy
losses, see Fig. \ref{fig:rel-beta} \cite{rel-beta}.

\begin{figure}[t]
 \centering
 {\centering\resizebox*{!}{8cm}{\includegraphics{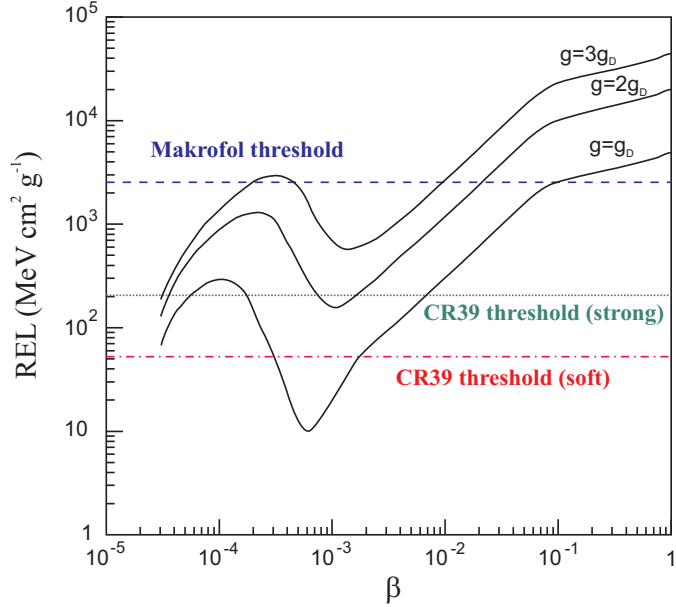}}}
\begin{quote}
 \caption{\small REL vs beta for magnetic monopoles with $g=g_D,~2g_D,~3g_D$.
The dashed lines represent the CR39 thresholds in soft and strong 
etching conditions and the Makrofol threshold (see Sect. \ref{sub:etching}).} 
\label{fig:rel-beta}
\end{quote}
 \end{figure}

 With the used etching conditions, the CR39 allows the detection of $(i)$ 
MMs with $g=g_D$ for $\beta \sim 10^{-4}$ and for $\beta > 10^{-2}$; $(ii)$
MMs with $g=2g_D$ for $\beta$ around $10^{-4}$ 
and for $\beta > 4 \cdot 10^{-3}$; $(iii)$ the whole $\beta$-range of 
$4 \cdot 10^{-5} < \beta < 1$ is accessible for MMs with $g > 2 g_D$ and 
for dyons. 

 For the Makrofol polycarbonate the detection threshold is at $Z/\beta \sim 50$
and REL $\sim 2.5$ GeV cm$^2$ g$^{-1}$ \cite{calibrations}; for this reason 
the use of Makrofol is restricted to  the search for fast MMs.

\begin{figure}[t]
 \centering
 {\centering\resizebox*{!}{6.2cm}{\includegraphics{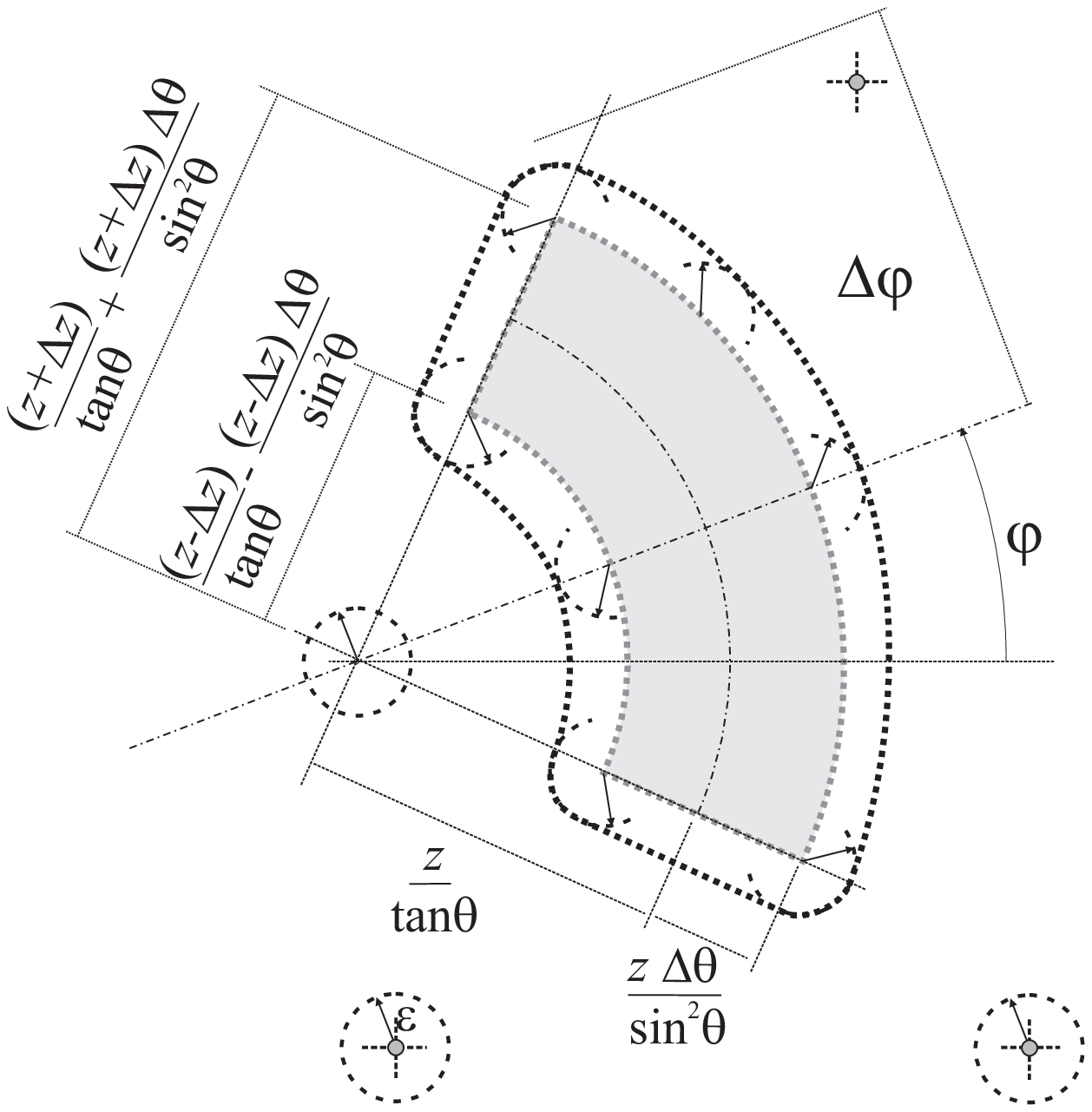}}}
 \hspace{1cm}
 {\centering\resizebox*{!}{6.2cm}{\includegraphics{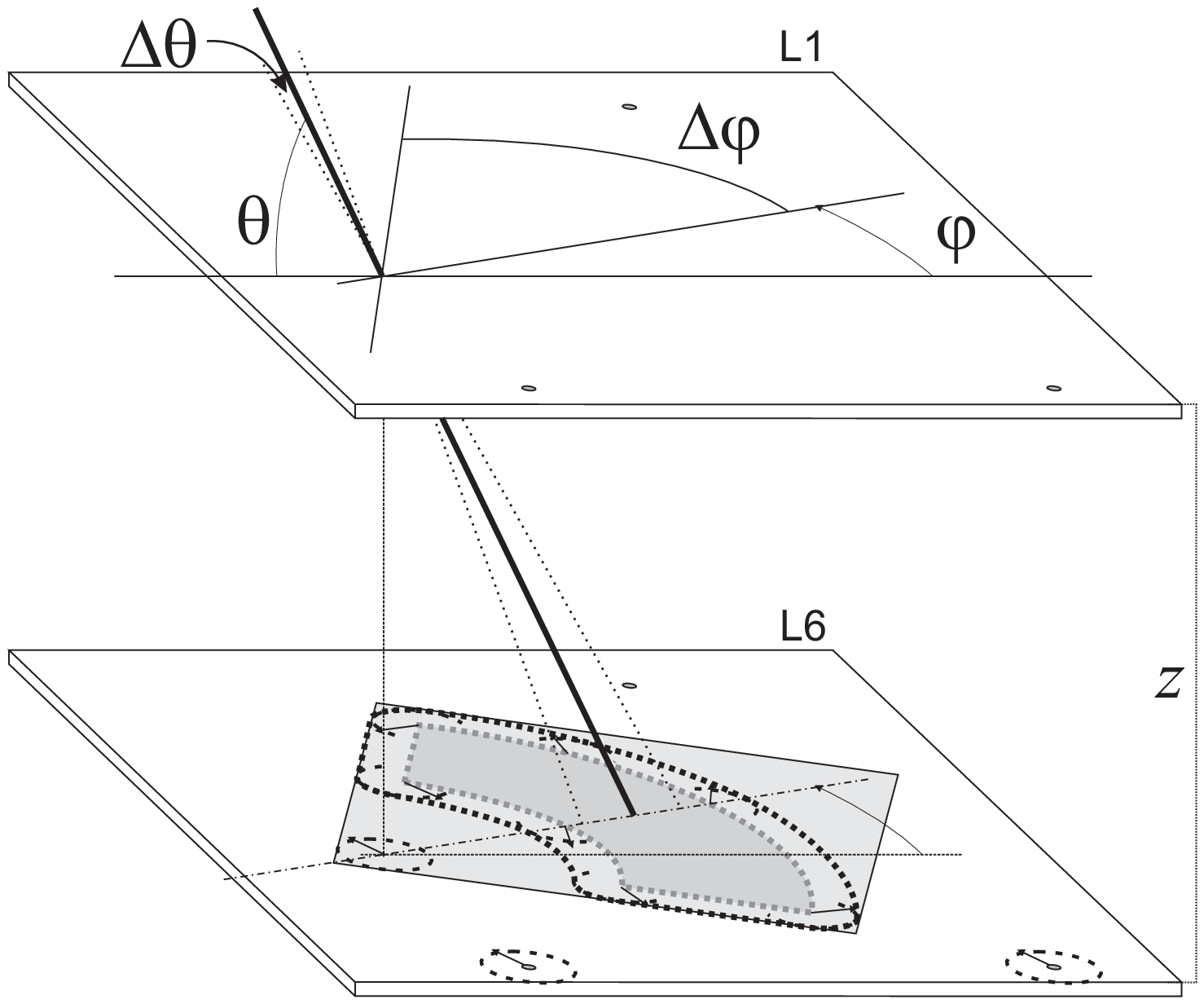}}}
\begin{quote}
 \caption{\small Illustration of the procedure used to define the 
``confidence'' area where the possible continuation of a candidate track 
inside two (or more) sheets of the same module was searched for (see text for
details).} 
\label{fig:illustration}
\end{quote}
 \end{figure}

\subsection{Analysis}
\label{sub:analysis}
After exposure at Chacaltaya the modules were brought back by air flights 
to Italy in order 
to be etched and analyzed in the Bologna laboratory. Three ``reference'' 
holes of 2 mm diameter were drilled in each module with a precision machine 
(the hole locations were defined to within 100 $\mu$m). This allowed us to 
follow the passage of a ``candidate''  through the stack.
The bags (envelopes) were opened, the detectors were labeled and their 
thicknesses were measured, using a micrometer, in 9 uniformly distributed 
points on the foil surface. 

The analysis of a SLIM module started by etching the uppermost CR39 sheet 
using strong conditions in order to reduce the CR39 thickness from 
1.4 mm to $\sim 0.9$ mm. After the strong etching, the CR39 sheet was scanned 
twice, with a stereo microscope, by different operators, 
 with a 3$\times$ magnification optical lens, looking for any possible 
correspondence of etch pits on the two opposite surfaces. The measured 
single scan efficiency was about 99\%; thus the double scan guarantees an 
efficiency of $\sim 100\%$ for finding a possible signal. 

Further observation of a ``suspicious correspondence'' was made with an 
optical $20 \div 40$$\times$ stereo microscope and classified either as a 
defect or a candidate track. This latter was then examined by an optical 
microscope with $6.3_{ob} \times 25_{oc}$ magnification and the axes of the 
base-cone ellipses in the front and back sides were measured. 
  
 A track was defined as  a ``candidate'' if the computed $p$ and incident 
angle $\theta$ on the front and back sides were equal to within 20\%. For 
each candidate the azimuth angle $\varphi$ and its 
position $P$ referred to the fiducial marks were also determined. 
The uncertainties $\Delta \theta$, $\Delta \varphi$ and $\Delta P$ 
defined a ``coincidence'' area ($< 0.5$ cm$^2$) 
around the candidate expected position in the other layers, as shown in 
Fig. \ref{fig:illustration}.

 In this case the lowermost CR39 layer was etched in soft etching 
conditions, and an accurate scan under an optical microscope with high 
magnification (500$\times$ or 1000$\times$) was 
performed in a square region around the candidate expected position, 
which included  the ``coincidence'' area. 
If a two-fold coincidence was detected, the CR39 middle layer was also 
analyzed. 

The bottom CR39 sheet was etched 
in about 50 cases; the third CR39 sheet was etched only in few cases, when 
there was still a possible uncertainty, and for checks ($\sim 16$ times). Some 
Makrofol foils were etched for reasons similar to the previous point and 
for other checks concerning the Makrofol itself ($\sim 12$ times).

\section{Results}
\label{sec:results}
From the detector calibration we computed the SLIM acceptance for 
downgoing IMMs with $g=g_D,~2g_D,~3g_D$ and for dyons.
  For the $i^{th}$ module of area $S_i$ the acceptance was computed as 
  
  \begin{equation}
  (S\Omega)_i = \pi S_i \left( 1 - \frac{1}{p^2} \right)
  \end{equation}
The total acceptance is the sum of all the individual contributions.

\begin{figure}[t]
\begin{center}
{\centering\resizebox*{!}{9cm}{\includegraphics{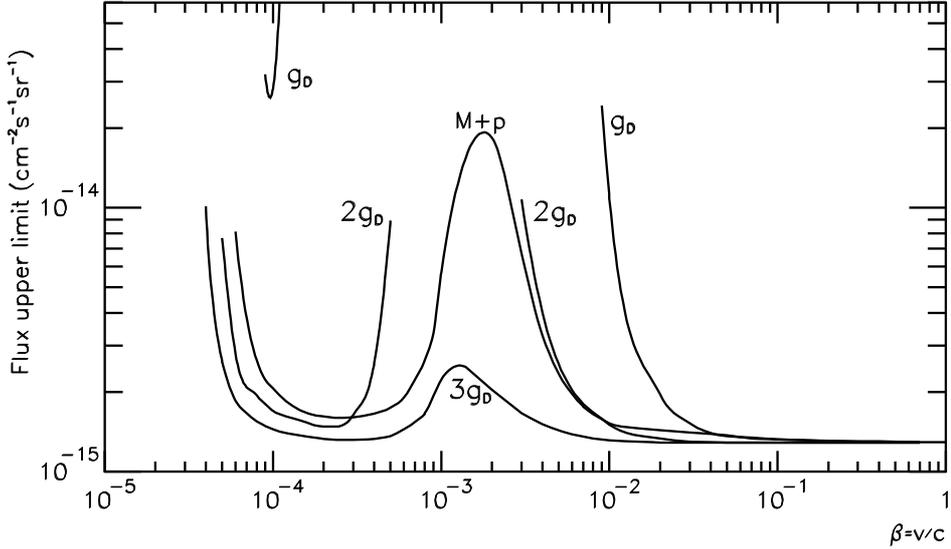}}}
\begin{quote} 
\caption{\small 90\% C.L. upper limits for a downgoing
 flux of IMMs with $g=g_D,~2g_D,~3g_D$ and for dyons (M+p , $g = g_D$) 
plotted vs $\beta$ (for strong etching). The poor limits at $\beta \sim 
10^{-3}$ arise because the REL is below the threshold (for $g_D$ and $2g_D$)
or slightly above the threshold (for $3g_D$ and dyons), see 
Sect. \ref{sub:cal}. }
\label {fig:limite}
\end{quote}
\end{center}
 \end{figure}

Since no candidates were found, the 90\% C.L. upper limit for a downgoing 
flux of IMMs and for dyons was computed as 
  
  \begin{equation}
  \phi = \frac{2.3}{(S\Omega) \cdot \Delta t \cdot \epsilon}
  \end{equation}
where $\Delta t$ is the mean exposure time (4.22 y), $S\Omega$ is the total 
acceptance, $\epsilon$ is the scanning efficiency estimated to be $\sim 1$.

The global 90\% C.L. upper limits for the flux of downgoing IMMs and dyons 
with velocities $\beta > 4 \cdot 10^{-5}$ were computed, as shown in Fig. 
\ref{fig:limite}. The flux limit for $\beta > 0.03$ is $\sim 1.3 \cdot 
10^{-15}$ cm$^{-2}$ s$^{-1}$ sr$^{-1}$.     
          
Two ``strange events'' were observed and were finally classified as 
manufacturing defects in a small subset of CR39 NTDs. These ``strange events'' 
are discussed in detail elsewhere \cite{balestra}.

 \begin{figure}[h!]
\centering
{\centering\resizebox*{!}{9cm}{\includegraphics{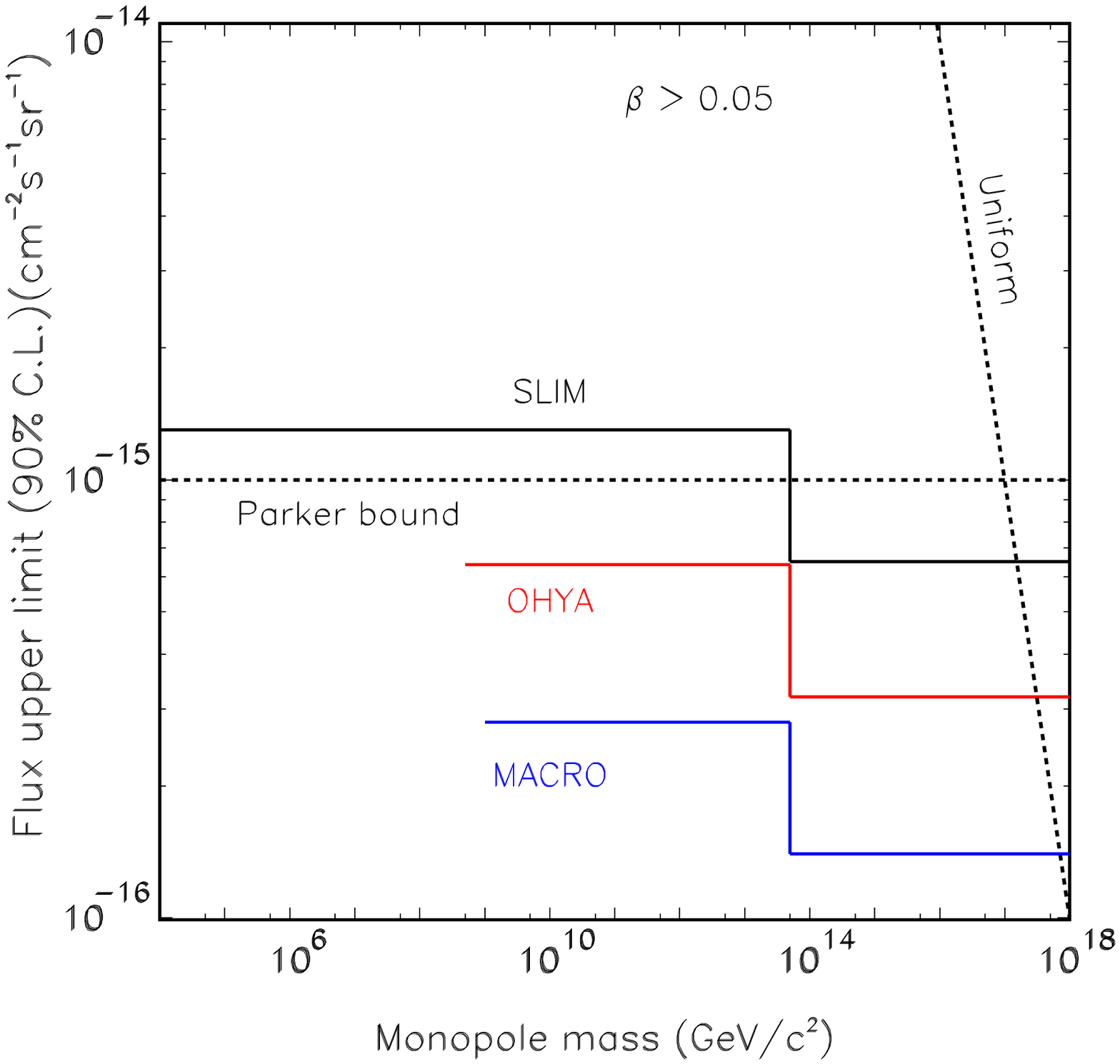}}}
\begin{quote}
 \caption{\small Flux upper limits for cosmic MMs of charge $g=g_D$ and 
$\beta > 0.05$ vs monopole mass. The figure shows the 90\% C.L. limits 
obtained by the SLIM,  MACRO \cite{MACRO} and OHYA \cite{orito} experiments. 
 MMs with masses smaller than $\sim 5 \cdot 10^{13}$ GeV are detected only 
if coming from above; MMs with masses larger than $\sim 5 \cdot 10^{13}$ GeV 
can traverse the Earth, so an isotropic flux is expected. The Parker 
bound \cite{parker}, obtained from the survival of the galactic 
magnetic field, and the 
limit obtained from the mass density for a uniform density of monopoles 
in the Universe \cite{uniform} are also plotted.} 
\label{fig:limi}
\end{quote}
 \end{figure}

\section{Conclusions}
\label{sec:conclu}
We etched and analyzed 427 m$^2$ of CR39, with an average exposure
time of 4.22 years. No candidate passed the search criteria. The 90\%
C.L. upper limits for a downgoing flux of fast ($\beta > 0.03$) 
IMM's coming from above
are at the level of $1.3 \cdot 10^{-15}$ cm$^{-2}$ sr$^{-1}$ s$^{-1}$. The 
complete $\beta$-dependence for MMs with $g=g_D,~2g_D,~3g_D$ and for dyons 
is shown in Fig. \ref{fig:limite}. 
  
 Superheavy GUT magnetic monopoles in the cosmic radiation can traverse 
the Earth. Therefore the SLIM limit on their flux is one half of the 
IMM flux: $\phi_{GUT} < 6.5 \cdot 10^{-16}$ cm$^{-2}$ s$^{-1}$ sr$^{-1}$ for 
$\beta > 0.03$ for $g=g_D$ \cite{MACRO}. 
 
 Fig. \ref{fig:limi} shows the flux upper limits for MMs of charge $g=g_D$ 
and $\beta > 0.05$ vs monopole mass. Note that the SLIM limit is $1.3 \cdot 
10^{-15}$ cm$^{-2}$ sr$^{-1}$ s$^{-1}$ for MM masses smaller than 
$\sim 5 \cdot 10^{13}$ GeV and $0.65 \cdot 10^{-15}$ cm$^{-2}$ sr$^{-1}$ 
s$^{-1}$ for masses larger than $\sim 5 \cdot 10^{13}$ GeV. In Fig. 
\ref{fig:limi} are also shown the limits obtained by the MACRO 
\cite{MACRO} and OHYA \cite{orito} experiments for $g=g_D$ magnetic monopoles 
with $\beta > 0.05$. 

SLIM is the first experiment to extend the cosmic radiation search for 
Magnetic Monopoles to masses lower than the GUT scale with a high sensitivity.
   
The addition of SLIM data to the MACRO data would improve the MACRO limits 
by only 18\%.

Large scale underwater and under ice neutrino telescopes (Amanda, IceCube, 
 ANTARES, NEMO) have the possibility to search for fast IMMs with $\beta > 0.5$
 to a level lower than the Parker bound \cite{2007,espi}.

\vspace{0.5cm}

{\Large \bf Acknowledgments}

We thank the engineering staff in BNL and in CERN for their help for the 
heavy ion calibration exposures. We acknowledge the collaboration of our 
technical staff, in particular L. Degli Esposti, G. Grandi and C. Valieri 
of INFN Bologna, and the technical staff of the Chacaltaya Laboratory. We 
thank A. Casoni for typing and correcting the manuscript. We thank INFN 
and ICTP for providing grants to non-Italian citizens. 

\newpage

\bibliographystyle{plain}

\begin{thebibliography}{99}
  
\bibitem{Dirac} P.A.M. Dirac, Proc. Roy. Soc. \textbf{133} (1931) 60.

\bibitem{biblio} G. Giacomelli et al., hep-ex/0005041. 

\bibitem{bertani}
M. Bertani et al., Europhys. Lett. \textbf{12} (1990) 613. \\
K. Kinoshita et al., Phys. Rev. D\textbf{46} (1992) R881. \\
G. Abbiendi et al., arXiv:0707.0404 [hep-ex].  \\
A. Abulencia et al., Phys. Rev. Lett. \textbf{96} (2006) 201801.\\
G.R. Kalbfleisch et al., Phys. Rev. Lett. \textbf{85} (2000) 5292. 

\bibitem{bakari}
D. Bakari et al., hep-ex/0004019.

\bibitem{MMs}
J. Preskill, Ann. Rev. Nucl. Part. Sci. \textbf{34} (1984) 461.\\
G. Giacomelli, Riv. Nuovo Cimento 7N\textbf{12} (1984) 1; Riv. Nuovo 
Cimento 16\textbf{N3} (1993) 1.\\
D.E. Groom, Phys. Rep. \textbf{140} (1986) 323. \\ 
G. Giacomelli et al, hep-ex/0702050. 

\bibitem{MACRO}
M. Ambrosio et al., Eur. Phys. J. C\textbf{25} (2002) 511;
Eur. Phys. J. C\textbf{26} (2002) 163; 
Astropart. Phys. \textbf{18} (2002) 27; 
Nucl. Instrum. Meth. A\textbf{486} (2002) 663.

\bibitem{IMMs}
G. Lazarides and Q. Shafi, Phys. Lett. B\textbf{148} (1984) 35. \\
P.H. Frampton and T.W. Kephart, {\it Phys. Rev.} D42 (1990) 3892. \\
T.W. Kephart and Q. Shafi, Phys. Lett. B\textbf{520} (2001) 313.

\bibitem{UHECR}
T.W. Kephart and T.J. Weiler, Astropart. Phys. \textbf{4} (1996) 271. \\
C.O. Escobar and R.A. Vasquez, Astropart. Phys. \textbf{10} (1999) 197.

\bibitem{proposal} 
D. Bakari et al., hep-ex/0003028. \\
S. Cecchini et al., Il Nuovo Cim. \textbf{24C} (2001) 639.

\bibitem{1998}
J. Derkaoui et al., Astropart. Phys. \textbf{10} (1999) 339; Astropart. 
Phys. \textbf{9} (1998) 173.

\bibitem{2007}
H. Wissing for the ICECUBE Coll., 30$^{th}$ ICRC (Merida, 2007), 
arXiv:0711.0353 [astro-ph], 139.

\bibitem{nuclr}
A. Witten, Phys. Rev. D\textbf{30} (1984) 272. \\ 
A. De Rujula and S. L. Glashow, Nature \textbf{312} (1984) 734. 

\bibitem{SLIM05/5}
S. Balestra et al., hep-ex/0506075; Czech. J. Phys. \textbf{56} (2006) 
A221; hep-ex/0601019.\\
S. Cecchini et al., Radiat. Meas. \textbf{40} (2005) 405. 

\bibitem{qballs}
S. Coleman, Nucl. Phys. B\textbf{262} (1985) 263. \\
A. Kusenko et al., Phys. Lett. B\textbf{418} (1998) 46.\\ 
J. Arafune et al., Phys. Rev. D\textbf{62} (2000) 105013.

\bibitem{SQM} S. Cecchini et al., arXiv:0805.1797 [hep-ex], submitted to
Eur. Phys. Jou. C. 

\bibitem{NTDsM}
S. Cecchini et al., Radiat. Meas. \textbf{34} (2001) 55.

\bibitem{calibrations}
S. Balestra et al., Nucl. Instrum. Meth. B\textbf{254} (2007) 254.\\
S. Manzoor et al., Radiat. Meas. \textbf{40} (2005) 433; Nucl. Phys. 
B Proc. Suppl. \textbf{172} (2007) 296.\\
G. Giacomelli et al., Nucl. Instrum. Meth. A\textbf{411} (1998) 41.

\bibitem{neutron}
H. Schraube et al., Rad. Prot. Dos. \textbf{84} (1999) 309. \\
A. Zanini et al., Il Nuovo Cim. \textbf{24C} (2001) 691. 

\bibitem{neutroni}
A. Zanini et al., Journ. Atm. and Solar-Terrestrial Phys. \textbf{67} 
(2005) 755.

\bibitem{fraggg}
S. Cecchini et al., arXiv:0801.3195 [nucl-ex], accepted by Nucl. Phys. A.

\bibitem{Elbeck}
A. Noll et al., Nucl. Tracks Radiat Meas. \textbf{15} (1988) 265.

\bibitem{rel-beta} J. Derkaoui et al., Astrop. Phys. \textbf{10} (1999) 339.

\bibitem{balestra}
S. Balestra et al., arXiv:0802.2056 [hep-ex].

\bibitem{orito} S. Orito et al., Phys. Rev. Lett. \textbf{66} (1991) 1951.

\bibitem{espi}
http://amanda.uci.edu/             ;             http://icecube.wisc.edu/ ;\\
http://antares.in2p3.fr/             ;             http://nemo.in2p3.fr/ .

\bibitem{parker} M.S. Turner et al., Phys. Rev. D\textbf{26} (1982) 1296.

\bibitem{uniform} A.H. Guth, Phys. Rev. D\textbf{23} (1981) 347.

\end{thebibliography}

\end{document}